\documentclass{mn2e}

\usepackage{epsfig}

\usepackage{mathptm}

\title{9C: A Survey of Radio Sources at 15~GHz with the Ryle Telescope }
\author[ ]
       { Elizabeth M.Waldram\thanks{E-mail: emw1@mrao.cam.ac.uk},  Guy G.Pooley,  Keith J.B.Grainge,  Michael E.Jones, 
\newauthor Richard D.E.Saunders, Paul F.Scott and  Angela C.Taylor \\
        Astrophysics Group, Cavendish Laboratory, Madingley Road, Cambridge, CB3 0HE}
\date{   }

\pagerange{\pageref{firstpage}--\pageref{lastpage}}
\pubyear{2002}

\begin{document}

\maketitle

\label{firstpage}

\begin{abstract}
The fields chosen for the first observations of the cosmic microwave background with the Very Small Array have been surveyed with the Ryle Telescope at 15 GHz. We have covered three regions around RA $00^{\rm h}20^{\rm m}$  Dec $+30^{\circ}$, RA $09^{\rm h}40^{\rm m}$ Dec $+32^{\circ}$ and RA $15^{\rm h}40^{\rm m}$ Dec $+43^{\circ}$ (J2000.0), an area of 520 deg$^{2}$.  There are 465 sources above the current completeness limit of $\approx25$ mJy, although a total of $\approx760$ sources have been detected, some as faint as 10 mJy.  This paper describes our techniques for observation and data analysis; it also includes source counts and some discussion of spectra and variability. Preliminary source lists are presented. 
\end{abstract}


\section{Introduction}
Foreground sources represent a major contaminant for cm-wave cosmic microwave background (CMB) measurements and our current survey with the Ryle Telescope at 15.2~GHz was set up as part of the observing strategy of the CMB telescope, the Very Small Array (VSA) (Watson et al. 2003). Its prime motivation has been to define a catalogue of the foreground sources which must be monitored by the VSA during its observations at a frequency of 34~GHz (Taylor et al. 2003). It is, however, of much wider interest, being the first survey covering an appreciable area at a radio frequency above the 4.8 GHz of the Green Bank survey (Gregory et al. 1996). In particular, it provides a means of identifying Gigahertz Peaked Spectrum (GPS) sources, which are important for the study of radio source evolution as well as being a significant foreground for CMB observations over a range of wavelengths. Since it will be a new, quite extensive survey, we have designated it `9C', or the 9th Cambridge survey.

The Ryle Telescope (RT) is an E-W radio synthesis telescope whose essential features are described by Jones (1991). The main problem in using the telescope for surveying the VSA fields is the small size of its primary beam envelope, which is only 6 arcmin \textsc{fwhm}, compared with 4\fdg6 for the VSA (for observations in its compact array). To overcome this, we have developed a new rastering technique which differs from the usual operation of the telescope in that we can cover a much wider area in a given time. A similar method was used to scan the foreground sources for the Cosmic Anisotropy Telescope (CAT) (Baker et al. 1999).

\section{Choice of Fields}
The fields for the first CMB observations with the VSA were chosen to be as free as possible from contamination by Galactic or extragalactic foreground radiation, whether synchrotron, free-free or from dust. An attempt was made to select fields likely to contain no very bright radio sources ($> 0.5 $ Jy) at 30~GHz. To make this prediction, sources in the Green Bank 4.8~GHz survey were matched with those in the NVSS 1.4~GHz survey (Condon et al. 1998), and their 30~GHz flux densities calculated simply from their spectral indices based on these two frequencies. This was known to be unreliable but, in the absence of a higher frequency survey, was the best indication available.

The first three areas were chosen to be more than $25^\circ$ from the Galactic plane, spaced in RA and observable from both Tenerife (for the VSA) and Cambridge, UK (for the RT). They are in the regions around RA $00^{\rm h}20^{\rm m}$  Dec $+30^\circ$, RA $09^{\rm h}40^{\rm m}$ Dec $+32^\circ$ and RA $15^{\rm h}40^{\rm m}$ Dec $+43^\circ$ (J2000.0). In each area a mosaic of overlapping VSA observations has been made and the RT survey is designed to cover the VSA primary beam out as far as a diameter of $9^\circ$ for each pointing in the mosaic. In practice, we have covered significantly wider areas than the minimum required (see section~6).

For more details of the criteria used in selecting the VSA fields see Taylor et al. 2003.
                         
\begin{figure}
        \centerline{\epsfig{file=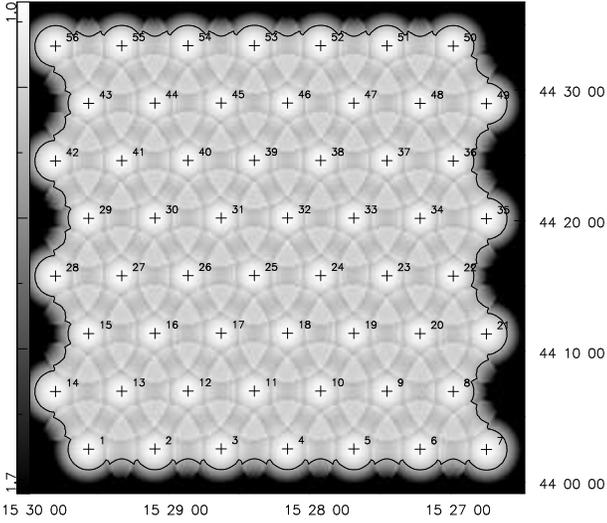,
        angle=270,
        width=8.0cm,clip=}}

        \caption{The sensitivity map corresponding to a $7\times8$ hexagonal array of pointing directions. The grey scale shows the value of $M_{\mathrm{n}}$ (see section 3.1). White indicates high sensitivity. (Here the coordinates are RA and Dec B1950.0.) }
\end{figure}   

\section{Observations and Data Analysis}
Since the area to be covered for each VSA pointing is  64 deg$^{2}$, while the area within the \textsc{fwhm} of the RT primary beam is only 0.01~deg$^{2}$, we have had to devise a technique whereby the RT survey can keep pace with the VSA observations. The principle of our method is to cover 1 deg$^{2}$ of sky in 12 hours by making a raster scan of 240 different pointing directions during that time. We use the E~-~W array of 5 aerials spaced to give a resolution of  25 arcsec, or, in some configurations of the telescope, 12 arcsec.  The 12 hours of data are combined into a single map which is used simply to identify peaks corresponding to possible radio sources. Each possible source is subsequently followed up with a short pointed observation of 10 - 15 minutes either to establish a reliable flux density or else to eliminate it as a false detection.

\begin{figure}
        \centerline{\epsfig{file=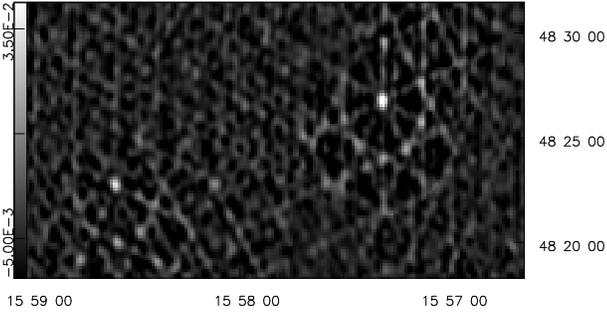,
        angle=270,
        width=8.0cm,clip=}}

        \caption{A section of a raster map with two sources of flux densities  50 mJy and  36 mJy. (Here the coordinates are RA and Dec B1950.0.)}
\end{figure}

\begin{figure}
        \centerline{\epsfig{file=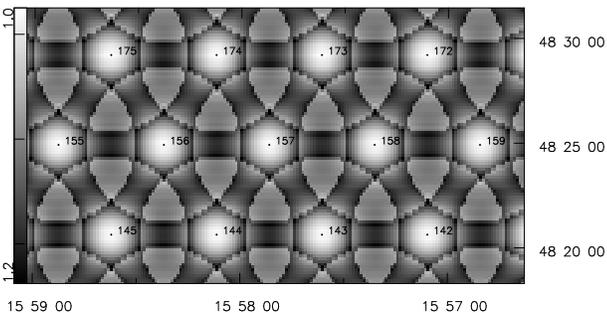,
        angle=270,
        width=8.0cm,clip=}}

        \caption{The section of the sensitivity map corresponding to the section of the raster map in Figure 2. The numbers show the sequence of pointing centres. The grey scale shows the value of $M_{\mathrm{n}}$ (see section 3.1). White indicates high sensitivity. }
\end{figure}

\subsection{The rastering technique}

The principle of our method is illustrated by one of our early trials. Figure 1 shows a small raster of 56 pointing centres in a $7\times8$ hexagonal array, spaced at intervals of 5 arcmin. In this example, there are 7 scans through the array during a 12-hour observation, in the sequence shown, with a dwell time on each centre at each pass of 12 of the 8-second data samples. A phase calibrator is observed periodically during the run. We make 56 small maps (so called `component' maps), one for each pointing, and, at this stage, with no correction for the primary beam envelope. The $uv$ aperture coverage is very sparse and is different for each map.

We calculate an appropriately weighted combination of the component maps to form a single large map with approximately uniform sensitivity (hereafter referred to as a `raster' map). The map value $M_{\mathrm{r}}$ at any point on the raster map is derived from the individual map values ($m_{i}$) and primary beam values ($p_{i}$) of up to 3 overlapping component maps. We assume that each component map has the same noise value $\sigma_{\mathrm{c}}$, and add the map values corrected for the primary beam, $\left(m_{i}/p_{i}\right)$,  weighted by $\left(p_{i}/\sigma_{\mathrm{c}}\right)^2$. This gives:

\[
M_{\mathrm{r}} = \left(\sum_{i=1}^{i_{\mathrm{max}}}m_{i}p_{i}\right)   \left(\sum_{i=1}^{i_{\mathrm{max}}}p_{i}^{2}\right)^{-1} \qquad \mathrm{where} \quad i_{\mathrm{max}} = 1, 2\ \mathrm{or}\ 3. 
\]
Only the region of each component map out as far as 0.3 of the primary beam maximum is used. We also calculate a second map in which the value $M_{\mathrm{n}}$ at the same point on the raster map is given by:
\[
M_{\mathrm{n}} = \left(\sum_{i=1}^{i_{\mathrm{max}}}p_{i}^{2}\right)^{-1/2} \qquad \mathrm{where} \quad i_\mathrm{max} = 1, 2\ \mathrm{or}\ 3.
\]
$M_{\mathrm{n}}$ is essentially the inverse of the sensitivity at any point on a raster map (see Figure 1). Thus, if in fact the noise values on the component maps were uniformly $\sigma_{\mathrm{c}}$, the noise at any point on the raster map would be $M_{\mathrm{n}}\sigma_{\mathrm{c}}$. We have chosen the spacing of the component pointing centres (i.e. 5 arcmin) such that $M_{\mathrm{n}}$ varies only over a range from 1 to a maximum vaue of $M^{\mathrm{max}}_{\mathrm{n}}$ within the main area of the raster map, where $M^{\mathrm{max}}_{\mathrm{n}}\approx 1.2$. $M_{\mathrm{n}}$ obviously rises steeply at the edges of the map and the contour in Figure 1 indicates the area outside which it exceeds $M^{\mathrm{max}}_{\mathrm{n}}$.

Although with this configuration of 56 pointing centres we can reach noise levels of 1.5 mJy, the area of sky covered in 12 hours is only  0.25 deg$^{2}$, and so in practice, for most of our observations, we use 240 pointing centres in a $15\times16$ hexagonal array, again spaced by 5 arcmin. There are then only 5 scans through the array during an observation, each with a theoretical dwell time of 4 of the 8-second data samples. The telescope pointing errors are monitored during the observations; for each baseline, those samples for which the mean pointing error of either aerial is excessive are excluded from the mapping process (we do not collect data during slewing). Consequently, one of the four 8-second data samples is usually lost. These larger raster maps cover an area of  1 deg$^{2}$ but the noise becomes about 4 mJy. Figure 2 shows a  section of such a map, with two sources of  50 mJy and 36 mJy, and Figure 3 the corresponding section of the sensitivity map. 

The aperture coverage is extremely sparse for these $15\times16$ rasters, as can be seen from the `spokes' in the source shapes. For example, the synthesized beam for the component map made from pointing centre 157 is  shown in Figure 4. However, our trials have shown that, since we are using the raster maps simply to locate peaks corresponding to possible sources, we can tolerate very sparse aperture coverage in order to maximise the speed of surveying. 

Our method is similar to `mosaicing' (Cornwell 1988, Sault et al. 1996) in that we are combining interferometric data from a series of pointing directions. However, mosaicing has traditionally been used to image extended objects which span several primary beams of the interferometer elements, whereas here we are only concerned to identify point sources. We do not attempt any joint deconvolution of the data, though we have shown that it is possible to apply CLEAN to the small component maps before combining them into a single raster map. We plan to CLEAN in the regions of very bright sources before publishing the final source catalogues but this has not been undertaken for the preliminary lists presented here. For the purpose of the analysis in this paper we have excluded the source 4C39.25 (9.5 Jy) and the area around it; it does not lie within the regions selected for our preliminary lists. (See section~6.)

\begin{figure}
        \centerline{\epsfig{file=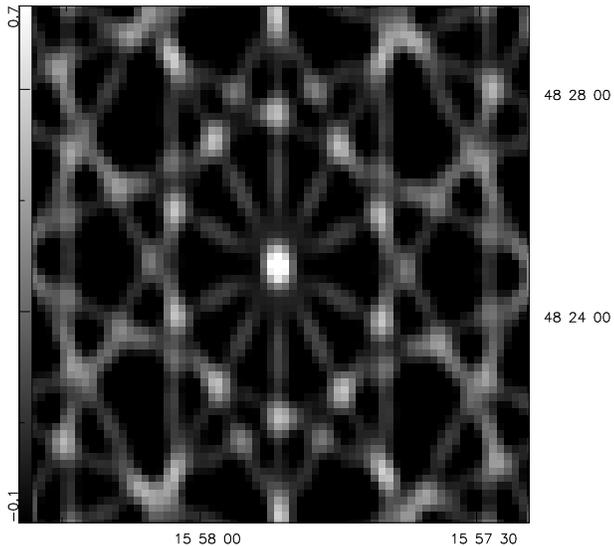,
        angle=270,
        width=8.0cm,clip=}}

        \caption{The synthesized beam for the component map corresponding to pointing centre 157 in Figure 3. (Here the coordinates are RA and Dec B1950.0) }
\end{figure}

\subsection{Source extraction}

Our source-finding method is constrained by the nature of the raster maps. Since the maps are not CLEANed, they include the effect of the strong sidelobes in the synthesized beam. Also, since different areas of the map are derived from different data, the noise may vary over the map, if there are varying weather conditions during the run. We therefore calculate both a mean noise value $\sigma$ for the whole map and an estimate of the local noise at points corresponding to the pointing centres, to check for very noisy areas; if necessary the observation is repeated.

Since the sensitivity of the raster map falls steeply at the edges, it is possible to use the map of $M_{\mathrm{n}}$ to define the area within which we search for sources, i.e. we can search the area corresponding to that within the $M^{\mathrm{max}}_{\mathrm{n}}$ contour. This has enabled us to use `scalloped' rather than straight edges for our search area and so to position the raster maps as close together as possible.  In our present method we scan this area of the raster map for local maximum pixels $\geq3\sigma$. Since the maps are sufficiently sampled, we can then, for each maximum, calculate a peak value, corresponding to a position interpolated between the grid points. (This is done by calculating the local map values on a successively finer grid, by repeated convolution with an appropriate gaussian-graded sinc function.) Peaks of height $\geq~5\sigma$ are selected to form a list of source candidates for subsequent pointed observations. 

This source extraction technique was developed in the course of the survey analysis. In particular, the values of $3\sigma$ for the pixel cut-off and $5\sigma$ for the peak cut-off were chosen after a number of trials; we have aimed to find as many weak sources as possible without accumulating an excess of spurious responses. However, initially we simply extracted from the raster maps the positions of local maximum pixels $\geq5\sigma$, and an area of $\sim100$ deg$^{2}$ out of the total 520 deg$^{2}$ has been processed in this way. To assess the effect on the catalogue, we have made a comparison of the results from our old and new algorithms over an area of $\sim 40$ deg$^{2}$. We find only 8 sources which were missed by the earlier method; all of these are below our estimated completeness limit of 25 mJy. Our conclusion is that, since we already have varying sensitivity over the whole survey area, the effect of the change in procedure is minimal.  

Pointed observations are made for each source candidate. A total observing time of about 15 min (much longer than the observing time for any one patch of the raster field) is sufficient to give a clear confirmation, or otherwise. By observing a group of nearby candidates in sequence, we spread this 15 min over two or three hours of hour angle and therefore are able to make useful maps to search for any extended sources; some 6\% are resolved in this way. We estimate that the flux densities have uncertainties of 5\% or less.

We find that the measured flux densities from the pointed observations are systematically higher (by about 10\%) than the estimates based on the raster maps. This has no significant consequence for the source list, since the rasters are simply used to locate the sources. The difference in scales is probably a result of the compromise necessary between keeping as much data as possible and discarding that with excessive pointing errors (see section 3.1).

\section{Completeness}

The layout of the observations is arranged to have small overlaps between the raster maps to ensure complete coverage of the required area. However, since the noise varies from raster map to raster map, as well as within the maps, there is some variation in sensitivity over the survey. We have investigated the completeness in two ways.

The first method has been to plot the ratio, $S_{\mathrm{r}}/S_{\mathrm{p}}$, of raster flux density to pointed flux density versus $S_{\mathrm{p}}$, as in Figure 5. It can be seen that, although for the brighter sources there is the expected scatter, below 25 mJy the ratio rises significantly. Our conclusion is that for these weaker sources the only peaks that are detected on the raster maps are those boosted by a positive local noise contribution, and that this implies a completeness limit of $\approx25$~mJy.

Secondly, we have made a deeper survey of a smaller area of  15 deg$^{2}$, within the area of the main survey, using mostly $9\times8$ raster scans. A 12 hour observation then covers an area of only 0.25 deg$^{2}$ but the noise level on a raster map becomes $\approx1.8$~mJy and a plot such as that in Figure 5 indicates that the completeness limit becomes $\approx10$~mJy. A comparison of the source count for the deeper area with that for the main area confirms our estimate of $\approx25$~mJy as the completeness limit of the main survey. It is not possible at present to be more precise about this completeness limit, since the deeper count includes too few sources, i.e. a total of only 68, with only 7 above 25 mJy. However, the indication is that at lower flux densities the completeness of the main survey is much reduced, falling to $\approx10$\% at the 10 mJy level.

\begin{figure}
        \centerline{\epsfig{file=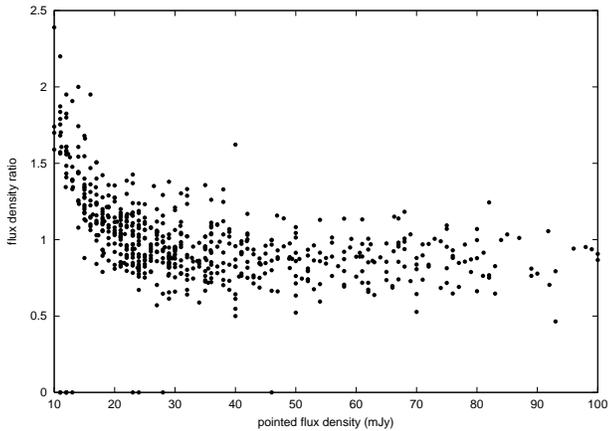,
        angle=270,
        width=8.0cm,clip=}}

        \caption{Plot of the ratio $(S_{\mathrm{r}}/S_{\mathrm{p}})$ of raster flux density to pointed flux density versus $S_{\mathrm{p}}$. (Where the ratio is shown as zero, the corresponding source lies in an area of a raster map which is hard to interpret. A pointed observation has been made but there is no definitive raster flux density.)} 
\end{figure}

\section{Some properties of the survey}

\subsection{Source Count}

\begin{figure}
        \centerline{\epsfig{file=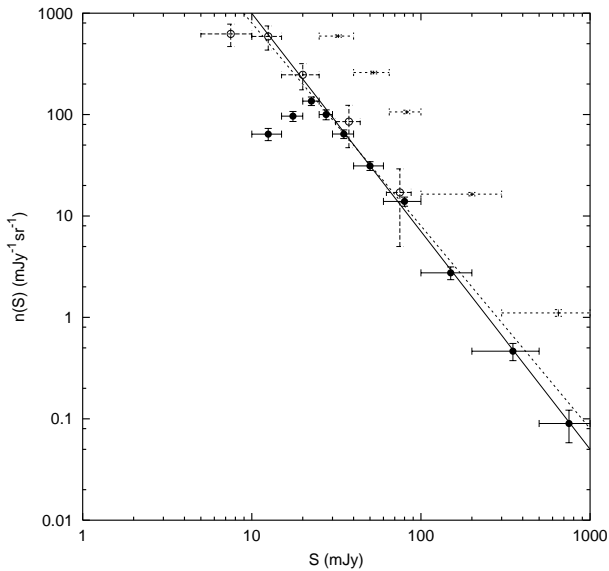,
        angle=270,
        width=8.0cm,clip=}}

        \caption{Plot of the differential source count for both the main survey (filled circles) and the deeper survey (open circles). The third count is the NVSS 1.4 GHz count, as calculated from our survey areas. The horizontal bars show the bin widths and the vertical bars the random errors. The solid line is the function we have fitted to the main count. The function from the earlier paper (Taylor et al. 2001) has been added (dashed) for comparison. (See section 5.1.)} 
\end{figure}

\begin{figure}
        \centerline{\epsfig{file=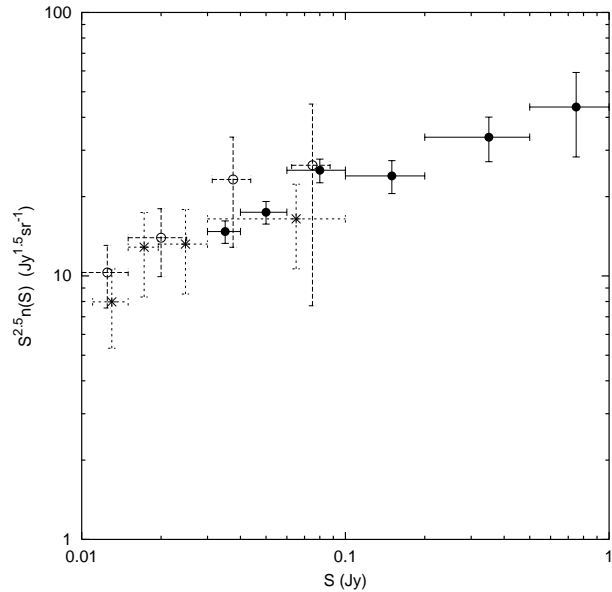,
        angle=270,
        width=8.0cm,clip=}}

        \caption{Plot of the normalised differential source count for both the main survey (filled circles) and the deeper survey (open circles). The third count (stars) is from an earlier survey at 15~GHz, made for the CAT observations. } 
\end{figure}

\begin{table}
\caption{Table showing the data used in fitting the source count for the main survey. The area is 0.1584 sr. $S$ is the flux density at the centre of the bin. We assume Poisson errors $\sqrt{N}$ in $N$.}
\begin{tabular}{ccc}
\hline
Bin centre & Width of bin & Number in bin \\
$S$ /Jy &  $W$ /Jy & $N$ \\
\hline

     0.0350  &    0.010     &      104 \\
     0.0500  &    0.020     &      103 \\
     0.0800  &    0.040     &       94 \\
     0.1500  &    0.100     &       49 \\
     0.3500  &    0.300     &       27 \\
     0.7500  &    0.500     &        8 \\

\hline
\end{tabular}
\end{table}

Figure 6 shows the differential source counts. It demonstrates the completeness limits, from the turnovers in the counts, at $\approx25$~mJy for the main survey and at $\approx10$~mJy for the deeper survey. For comparison, we have included the 1.4~GHz NVSS differential count, as calculated from the sources in that catalogue in our survey areas. 

Figure 7 shows the normalised differential counts $S^{2.5}n(S)$, i.e. our counts divided by the counts expected in a Euclidean universe. Points considered to be affected by incompleteness have been omitted; these include the point in the main survey corresponding to the 25-30 mJy bin, which may be marginally affected as our completeness estimate is only approximate. We also show a count from an earlier survey at 15~GHz that we made over an area of 17 deg$^{2}$ during the source scanning for the CAT observations. This was in a completely different area of sky from the current survey region. However, the deeper survey described here was made over an area within the main survey and therefore the two contain some sources in common.

We have calculated a least-squares fit of the function $ \mathrm{A} S^{\mathrm{b}}$ to the differential count of the main survey, using the data in Table 1 , with weights appropriate to the Poisson errors $\sqrt{N}$ in $N$,  and find:

\[
n(S) \equiv \frac{{\rm d}N}{{\rm d}S} \approx 51 \left( \frac{S}{\rm Jy} \right)^{-2.15}
\, {\rm Jy}^{-1}{\rm sr}^{-1}
\]

The effect of random errors in the flux densities is to produce an overestimate of the value of $N$ in each bin. However, in our case this is negligible, since the 5\% uncertainty would cause only $\sim~0.75$ \% increase.
We have, however, made a correction for the bin widths. We take $S$ to be the value at the centre of a bin and multiply the corresponding $N$  by a factor $(1 - r^{2})$ where $r = W/2S$. Here we are making the approximation that $n(S) \propto S^{-2}$ within any bin but are making no assumption about the size of the bin. 

The survey is entirely limited by noise, and not by source confusion. At 10 mJy, there are some 7200 sources per steradian, or $6.10^6$~$\mathrm{arcsec}^2$ per source. The beamwidth of the observations described here, $25 \times 25\mathrm{cosec}\delta$ $\mathrm{arcsec}^2$, means that there are typically 5000 beam areas per source, far above the value at which early radio source counts suffered from `confusion'. Since the pointed observations are used to make adequate maps of each source, there is very seldom any problem in deciding whether a multiple source is really that or a chance coincidence. 

Our fitted function is consistent with the data presented in the earlier source count paper (Taylor et al. 2001) but should provide a more accurate representation of the count, since it is based on a much larger sample of sources. The function quoted in that paper has been added to Figure 6 for comparison.

\begin{figure}
        \centerline{\epsfig{file=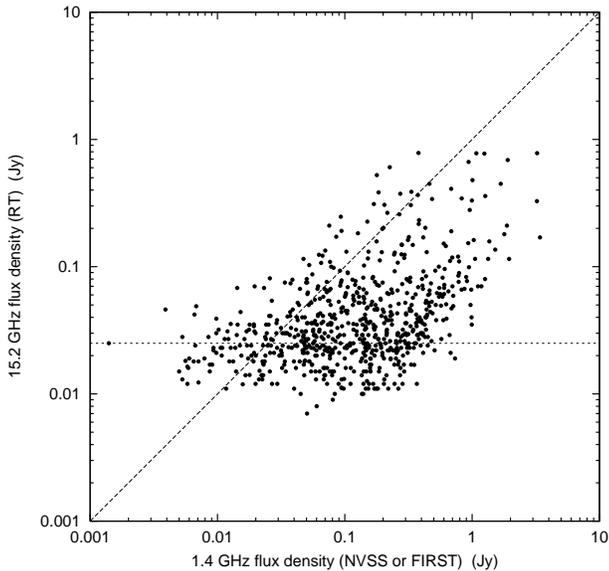,
        angle=270,
        width=8.0cm,clip=}}

        \caption{Plot of RT 15.2~GHz pointed flux densities (in the main survey) v. NVSS (or FIRST) 1.4~GHz flux densities, with a line indicating zero spectral index.  The horizontal line is the estimated completeness limit.}

\end{figure}

\begin{figure}
        \centerline{\epsfig{file=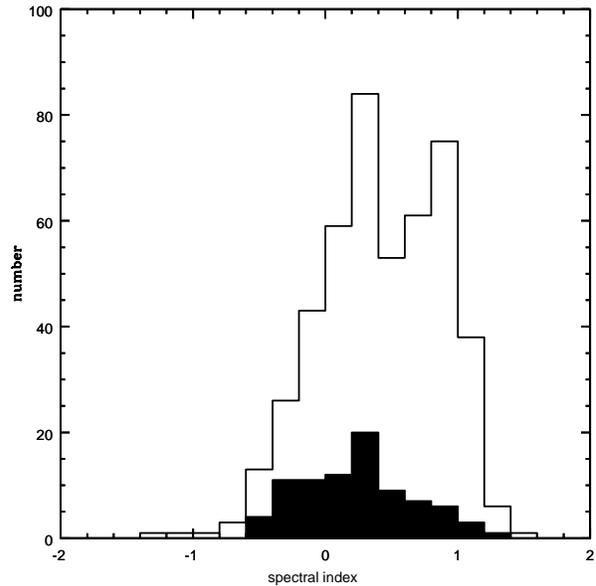,
        width=8.0cm,clip=}}
       \caption{Histogram (unshaded) of the spectral index distribution $\alpha_{1.4}^{15.2}$ for the 465 sources $\geq25$ mJy in our survey, where $\alpha$ is defined by $S\propto \nu^{-\alpha}$ . The shaded histogram shows the spectral index distribution for the 84 sources $\geq100$ mJy. }
\end{figure}

\begin{figure}
        \centerline{\epsfig{file=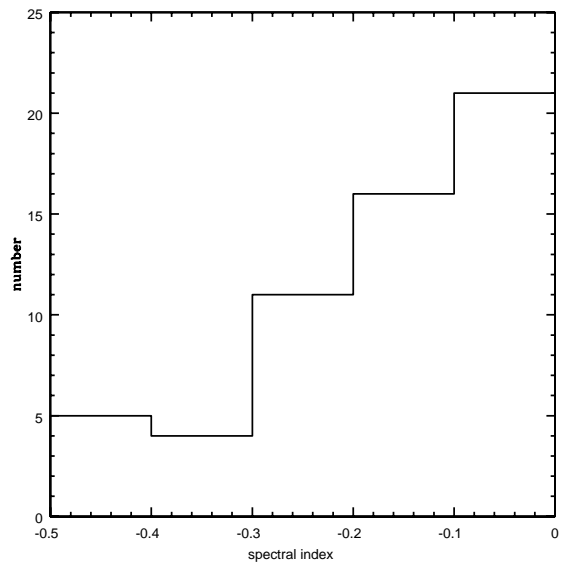,
        width=8.0cm,clip=}}
        \caption{Histogram of the spectral index distribution $\alpha_{1.4}^{15.2}$ for the sources $\geq25$ mJy in both our survey and NVSS, with $\alpha< 0$ .}
\end{figure}

\subsection{Correlation with NVSS at 1.4~GHz}
We have correlated our source lists with the catalogue from the 1.4~GHz NVSS survey which has a similar resolution to ours (see Figure 8). All the sources in the main survey in this paper have counterparts in NVSS, apart from two, one of which has been found in the deeper 1.4~GHz FIRST survey (Becker et al. 1995), leaving only one hitherto uncatalogued source. The horizontal line in Figure 8 marks the estimated completeness limit of 25 mJy; there are 465 sources above this limit.

A histogram of the spectral index distribution $\alpha_{1.4}^{15.2}$ for the 465 sources above the survey limit of 25 mJy is shown in Figure 9, where $\alpha$ is defined by $S\propto \nu^{-\alpha}$. Of these sources, 88 (19\%) have inverted spectra, i.e. $\alpha_{1.4}^{15.2} < 0 $. However, the distribution of spectral index for the 84 sources $\geq100$~mJy (shaded in Figure 9) appears to be significantly different, with 26~(31\%) having $\alpha_{1.4}^{15.2} < 0 $. 

Our correlation with NVSS shows the importance of the new survey in the investigation of the source population at high frequencies. This can be illustrated by considering all the sources $\geq25$~mJy in NVSS within our areas, 4089 sources in all. A possible procedure might have been to make pointed observations of all these sources at 15~GHz, in which case we should have found 434 sources above our survey limit of 25 mJy. However, 31 sources above our limit (i.e. 6.7 \% of the total) would have been missed, since their flux densities in NVSS are $< 25$ mJy.

Our correlation can also give some indication of the proportion of sources in
the NVSS catalogue with inverted spectra; this is important for any attempt to predict the source population at high frequencies from that at 1.4 GHz.  Out of the 4089 sources $\geq25$~mJy in NVSS in our areas, we find 57, above our survey limit of 25~mJy, to have $\alpha_{1.4}^{15.2} < 0 $, i.e. 1.4\% have rising spectra between 1.4 and 15.2~GHz. We clearly cannot plot the complete spectral index distribution for the 4089 sources but the distribution of these negative spectral indices is shown in Figure 10. This cannot, of course, provide information about inverted spectrum sources $< 25$~mJy in NVSS, though we can see from Figure~9 that some do have more extreme negative spectral indices.  

We should emphasize that the observations at 1.4 and 15.2~GHz are not simultaneous and many of the sources are likely to be highly variable.

\begin{figure}
        \centerline{\epsfig{file=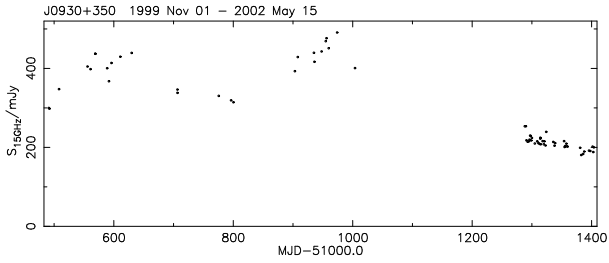,
        angle=270,
        width=8.0cm,clip=}}

        \caption{The variable source J0930+350: November 1999 to May 2002}

\end{figure}

\begin{figure}
        \centerline{\epsfig{file=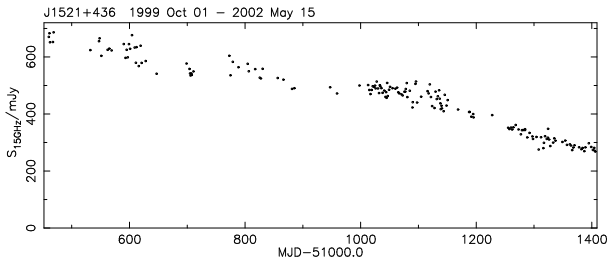,
        angle=270,
        width=8.0cm,clip=}}

        \caption{The variable source J1521+436: October 1999 to May 2002}

\end{figure}

\begin{figure}
        \centerline{\epsfig{file=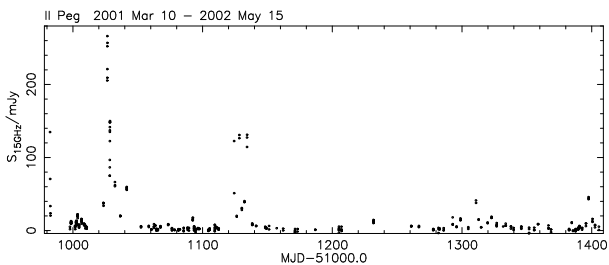,
         angle=270,
         width=8.0cm,clip=}}

        \caption{The variable RS CVn star II Pegasi: March 2001 to May 2002}

\end{figure}

\subsection{Variability}
We have made a preliminary check on variability by re-observing a sample of all sources above 60 mJy in an area of 94 deg$^{2}$. The time intervals were between 1 and 14 months. Out of the 35 sources, 8 were found to have varied by more than 20\% and, of these, 1 by more than 30\% and 1 by more than 60\%. Examples of two highly variable sources, used as phase calibrators in our survey, are shown in Figures 11 and 12. The estimated uncertainties in the flux densities are $\leq5$\%.

We have also identified the highly variable RS CVn star II~Pegasi (van den Oord and de Bruyn 1994) which lies within our survey area (see Figure 13). As in the detection by van den Oord and de Bruyn, we were alerted by the variation in flux density during the mapping observation. (This source has been excluded from the source lists used in this paper.)   

\section{Preliminary source lists}

\subsection{Sky coverage}
This is an on-going survey and, as explained earlier, its primary aim has been to cover the fields of the VSA observations, each of which contains a mosaic of pointing directions. This means that the regions of sky corresponding to our current source lists are necessarily ragged in shape and difficult to specify concisely. We have been working on extending the survey to form more easily defined regions and shall publish these in due course. For the purpose of this paper, in order to be able to define the areas simply and precisely, we have selected three circular areas, designated VSA1, VSA2, VSA3 (see Table 2), and present the source lists corresponding to these (Tables 4,5,6).

\subsection{Description of the lists}
We have included in these lists only those sources with flux densities above our estimated completeness limit of 25 mJy (see section 4), a total of 242 sources.The definitions of the entries are shown in Table 3. 

The position of a source is derived from the raster map, unless the follow-up pointed observation indicates a significant difference, in which case that position is substituted. We have checked the positions of a sample of 19 sources which appear on two raster maps and estimate that our position accuracy is better than 10 arcsec. For bright sources, the accuracy is better than 3 arcsec, as determined from the observations of the 17 sources in our VSA2 list, all brighter than 60 mJy at 15 GHz, which are found in the Jodrell-VLA calibrator survey ( Wilkinson et al. 1998 and references therein).

Our flux densities are those from the pointed observations. The uncertainty in these is dominated by the random uncertainty in the flux calibration, which we estimate to be $\sim 5$ \%. Since many of the sources are highly variable, we also quote the date of the pointed observation.

\begin{table}
\caption{The circular areas corresponding to the source lists}
\begin{tabular}{ccccc}
\hline
&  Centre J2000 &  Centre B1950 &  Radius  & Area   \\
&  RA $^{\rm h}\ ^{\rm m}\ ^{\rm s}$ Dec $^\circ\ ^\prime\ ^{\prime\prime}$  & RA $^{\rm h}\ ^{\rm m}\ ^{\rm s}$ Dec $^\circ\ ^\prime\ ^{\prime\prime}$ & deg   & deg$^{2}$ \\
\hline
VSA1 & $00\ 17\ 36.5\ \ 30\ 16\ 39$ & $00\ 15\ 00.0\ \ 30\ 00\ 00$ & 5.5 & 95.0  \\
VSA2 & $09\ 40\ 57.7\ \ 31\ 46\ 21$ & $09\ 38\ 00.0\ \ 32\ 00\ 00$ & 6.0 & 113.0 \\
VSA3 & $15\ 36\ 42.7\ \ 43\ 20\ 11$ & $15\ 35\ 00.0\ \ 43\ 30\ 00$ & 5.0 & 78.5  \\
\hline
\end{tabular}
\end{table}

\begin{table}
\caption{Definitions of the entries in the source lists}
\begin{tabular}{cll}
\hline
Column & Description & Example   \\
\hline
1 & Source name & 9CJ0002+3032 \\
2 & J2000 RA $^{\rm h}\ ^{\rm m}\ ^{\rm s}$ & $00\ 02\ 49.2$ \\
3 & J2000 Dec $^\circ\ ^\prime\ ^{\prime\prime}$ &  $30\ 32\ 43$ \\
4 & B1950 RA $^{\rm h}\ ^{\rm m}\ ^{\rm s}$ & $00\ 00\ 15.2$ \\
5 & B1950 Dec $^\circ\ ^\prime\ ^{\prime\prime}$ &$ 30\ 16\ 01$ \\
6 & Flux density  Jy & 0.046 \\
7 & Date yymmdd & 000531 \\
\hline
\end{tabular}
\end{table} 

\subsection{Availability of the catalogues}
These lists are available via our web page \textit{http://www.mrao.cam.ac.uk/surveys/index.html}. They are preliminary lists only and will be extended and refined as the survey proceeds. The web page will be updated accordingly.

\section{Conclusions and further work}
Our rastering technique with the Ryle Telescope has been used successfully to produce the first section of the 9C survey at 15~GHz; hitherto there has been no comparable high frequency radio survey of any extent. Although it has been designed specifically for use with the cm-wave CMB observations of the Very Small Array, our survey has much wider implications. We have explored several of its properties: the source count, the correlation with the NVSS 1.4~GHz survey, and the variability of the sources. These are important in studies of the radio source population at high frequencies, as well as in predicting the contaminating effect of foreground sources on CMB observations over a range of wavelengths. 

The completeness limit of the main part of the survey described here is $\approx25$ mJy but we are currently surveying further areas at the deeper level in order to provide appropriate source lists for the latest VSA observations, which are being made with its extended array at an increased sensitivity. These areas of the survey should reach a completeness limit of $\approx10$ mJy.

\section*{Acknowledgments}
We are grateful to the staff of our observatory for the operation of the Ryle Telescope, which is funded by PPARC. ACT acknowledges a PPARC studentship.

\begin{table*}
\caption{Source list for the VSA1 field}
\begin{tabular}{ccccccccccc}
Source name  &&    RA J2000 &  Dec J2000    &&    RA B1950 & Dec B1950    &&   Flux &&  Date  \\
&&&&&&&&density (Jy) \\
\hline
\\
9CJ0000+2914 && 00 00 35.2     & 29 14 28     && 23 58 01.5     & 28 57 47     && 0.036 && 010119 \\
9CJ0002+3139 && 00 02 14.2     & 31 39 42     && 23 59 40.2     & 31 23 00     && 0.036 && 000622 \\
9CJ0002+3032 && 00 02 49.2     & 30 32 43     && 00 00 15.2     & 30 16 01     && 0.046 && 000531 \\
9CJ0002+2942 && 00 02 52.4     & 29 42 55     && 00 00 18.4     & 29 26 13     && 0.052 && 000819 \\
9CJ0003+2740 && 00 03 13.0     & 27 40 46     && 00 00 39.0     & 27 24 04     && 0.075 && 000619 \\
9CJ0003+3010 && 00 03 55.6     & 30 10 02     && 00 01 21.3     & 29 53 20     && 0.081 && 000531 \\
9CJ0003+3241 && 00 03 56.3     & 32 41 54     && 00 01 22.0     & 32 25 12     && 0.065 && 000622 \\
9CJ0004+2637 && 00 04 23.7     & 26 37 52     && 00 01 49.5     & 26 21 10     && 0.071 && 000531 \\
9CJ0004+2946 && 00 04 34.6     & 29 46 17     && 00 02 00.2     & 29 29 35     && 0.067 && 000531 \\
9CJ0005+3206 && 00 05 40.4     & 32 06 06     && 00 03 05.9     & 31 49 25     && 0.096 && 000622 \\
9CJ0005+3139 && 00 05 55.8     & 31 39 49     && 00 03 21.2     & 31 23 07     && 0.076 && 000622 \\
9CJ0006+3422 && 00 06 07.3     & 34 22 19     && 00 03 32.6     & 34 05 38     && 0.075 && 000622 \\
9CJ0006+2523 && 00 06 28.5     & 25 23 01     && 00 03 54.0     & 25 06 19     && 0.050 && 001107 \\
9CJ0010+3403 && 00 10 03.8     & 34 03 12     && 00 07 28.3     & 33 46 31     && 0.029 && 000321 \\
9CJ0010+2838 && 00 10 11.1     & 28 38 11     && 00 07 35.9     & 28 21 30     && 0.061 && 000904 \\
9CJ0010+2854 && 00 10 27.8     & 28 54 58     && 00 07 52.5     & 28 38 16     && 0.038 && 000904 \\
9CJ0010+2717 && 00 10 28.7     & 27 17 54     && 00 07 53.6     & 27 01 13     && 0.036 && 000405 \\
9CJ0010+2619 && 00 10 36.3     & 26 19 19     && 00 08 01.2     & 26 02 38     && 0.065 && 000405 \\
9CJ0010+2956 && 00 10 42.5     & 29 56 12     && 00 08 07.1     & 29 39 31     && 0.061 && 000325 \\
9CJ0010+2650 && 00 10 51.3     & 26 50 28     && 00 08 16.1     & 26 33 47     && 0.036 && 000405 \\
9CJ0011+3443 && 00 11 17.3     & 34 43 34     && 00 08 41.5     & 34 26 53     && 0.029 && 001011 \\
9CJ0011+2803 && 00 11 34.2     & 28 03 48     && 00 08 58.8     & 27 47 07     && 0.039 && 000405 \\
9CJ0011+2928 && 00 11 46.0     & 29 28 30     && 00 09 10.5     & 29 11 49     && 0.053 && 000325 \\
9CJ0012+2702 && 00 12 38.2     & 27 02 40     && 00 10 02.7     & 26 45 59     && 0.072 && 000816 \\
9CJ0012+3353 && 00 12 47.3     & 33 53 36     && 00 10 11.3     & 33 36 55     && 0.098 && 000221 \\
9CJ0013+2834 && 00 13 32.7     & 28 34 53     && 00 10 57.0     & 28 18 12     && 0.028 && 000108 \\
9CJ0013+3441 && 00 13 44.1     & 34 41 41     && 00 11 07.8     & 34 25 01     && 0.136 && 001005 \\
9CJ0013+2646 && 00 13 46.5     & 26 46 42     && 00 11 10.9     & 26 30 02     && 0.030 && 000816 \\
9CJ0014+2815 && 00 14 33.8     & 28 15 07     && 00 11 58.0     & 27 58 27     && 0.048 && 000108 \\
9CJ0015+3216 && 00 15 06.2     & 32 16 13     && 00 12 29.9     & 31 59 33     && 0.448 && 000107 \\
9CJ0015+3052 && 00 15 36.1     & 30 52 24     && 00 12 59.9     & 30 35 44     && 0.035 && 000108 \\
9CJ0016+2510 && 00 16 39.8     & 25 10 28     && 00 14 04.0     & 24 53 49     && 0.052 && 000816 \\
9CJ0018+2921 && 00 18 12.5     & 29 21 24     && 00 15 35.9     & 29 04 45     && 0.084 && 000109 \\
9CJ0019+2817 && 00 19 08.9     & 28 17 56     && 00 16 32.3     & 28 01 17     && 0.035 && 000001 \\
9CJ0019+2956 && 00 19 37.8     & 29 56 02     && 00 17 01.0     & 29 39 23     && 0.043 && 000109 \\
9CJ0019+2602 && 00 19 39.8     & 26 02 52     && 00 17 03.4     & 25 46 14     && 0.409 && 000816 \\
9CJ0019+2647 && 00 19 52.5     & 26 47 31     && 00 17 16.0     & 26 30 52     && 0.053 && 000531 \\
9CJ0019+3320 && 00 19 58.6     & 33 20 02     && 00 17 21.3     & 33 03 24     && 0.036 && 000110 \\
9CJ0020+3152 && 00 20 50.6     & 31 52 30     && 00 18 13.3     & 31 35 51     && 0.041 && 000109 \\
9CJ0021+2711 && 00 21 28.7     & 27 11 44     && 00 18 52.0     & 26 55 05     && 0.039 && 000221 \\
9CJ0021+3226 && 00 21 29.9     & 32 26 57     && 00 18 52.5     & 32 10 19     && 0.027 && 000109 \\
9CJ0022+3250 && 00 22 43.8     & 32 50 46     && 00 20 06.0     & 32 34 09     && 0.025 && 000110 \\
9CJ0023+3114 && 00 23 10.0     & 31 14 01     && 00 20 32.4     & 30 57 24     && 0.033 && 000110 \\
9CJ0023+2734 && 00 23 14.5     & 27 34 30     && 00 20 37.4     & 27 17 52     && 0.067 && 000221 \\
9CJ0023+2539 && 00 23 23.1     & 25 39 19     && 00 20 46.2     & 25 22 42     && 0.124 && 001006 \\
9CJ0024+2510 && 00 24 16.0     & 25 10 38     && 00 21 39.1     & 24 54 01     && 0.047 && 001006 \\
9CJ0026+3508 && 00 26 41.7     & 35 08 42     && 00 24 02.8     & 34 52 07     && 0.152 && 000911 \\
9CJ0027+2830 && 00 27 00.6     & 28 30 23     && 00 24 22.8     & 28 13 47     && 0.026 && 000110 \\
9CJ0027+2607 && 00 27 27.0     & 26 07 08     && 00 24 49.5     & 25 50 33     && 0.032 && 000601 \\
9CJ0027+2555 && 00 27 29.6     & 25 55 32     && 00 24 52.2     & 25 38 57     && 0.052 && 001006 \\
9CJ0028+3103 && 00 28 10.8     & 31 03 50     && 00 25 32.4     & 30 47 15     && 0.027 && 000303 \\
9CJ0028+2914 && 00 28 17.1     & 29 14 29     && 00 25 39.0     & 28 57 54     && 0.089 && 991127 \\
9CJ0028+2954 && 00 28 31.9     & 29 54 52     && 00 25 53.6     & 29 38 17     && 0.030 && 001006 \\
9CJ0029+3456 && 00 29 14.3     & 34 56 34     && 00 26 34.9     & 34 39 59     && 0.689 && 001007 \\
9CJ0029+3244 && 00 29 33.0     & 32 44 52     && 00 26 54.0     & 32 28 17     && 0.047 && 000305 \\
9CJ0030+2957 && 00 30 05.3     & 29 57 07     && 00 27 26.8     & 29 40 33     && 0.032 && 991127 \\
9CJ0030+3415 && 00 30 21.5     & 34 15 58     && 00 27 42.0     & 33 59 24     && 0.029 && 000305 \\
9CJ0030+2833 && 00 30 28.7     & 28 33 37     && 00 27 50.4     & 28 17 04     && 0.049 && 000110 \\
9CJ0030+2809 && 00 30 57.1     & 28 09 43     && 00 28 18.8     & 27 53 10     && 0.030 && 000110 \\
9CJ0031+3016 && 00 31 21.9     & 30 16 04     && 00 28 43.1     & 29 59 31     && 0.056 && 000303 \\

\end{tabular}
\end{table*}

\begin{table*}
\contcaption{}
\begin{tabular}{ccccccccccc}
Source name  &&    RA J2000 &  Dec J2000    &&    RA B1950 & Dec B1950    &&   Flux &&  Date  \\
&&&&&&&&density (Jy) \\
\hline
\\
9CJ0032+2758 && 00 32 44.6     & 27 58 56     && 00 30 06.0     & 27 42 24     && 0.038 && 000222 \\
9CJ0033+2752 && 00 33 09.7     & 27 52 30     && 00 30 31.1     & 27 35 58     && 0.026 && 000308 \\
9CJ0034+2754 && 00 34 43.5     & 27 54 26     && 00 32 04.6     & 27 37 54     && 0.344 && 000303 \\
9CJ0036+3151 && 00 36 48.1     & 31 51 14     && 00 34 07.9     & 31 34 44     && 0.080 && 000309 \\
9CJ0036+2958 && 00 36 59.8     & 29 58 58     && 00 34 20.1     & 29 42 28     && 0.122 && 000305 \\
9CJ0037+3002 && 00 37 10.2     & 30 02 44     && 00 34 30.4     & 29 46 14     && 0.034 && 000817 \\
9CJ0038+2932 && 00 38 31.4     & 29 32 19     && 00 35 51.5     & 29 15 50     && 0.030 && 000305 \\
9CJ0039+3114 && 00 39 08.4     & 31 14 21     && 00 36 28.0     & 30 57 52     && 0.026 && 000610 \\
9CJ0041+3211 && 00 41 16.0     & 32 11 08     && 00 38 35.0     & 31 54 41     && 0.170 && 000604 \\
9CJ0042+3005 && 00 42 39.8     & 30 05 59     && 00 39 59.2     & 29 49 34     && 0.029 && 000604 \\
9CJ2352+3030 && 23 52 54.7     & 30 30 24     && 23 50 22.3     & 30 13 43     && 0.036 && 010130 \\
9CJ2353+3040 && 23 53 06.7     & 30 40 49     && 23 50 34.3     & 30 24 08     && 0.037 && 010130 \\
9CJ2353+3136 && 23 53 19.5     & 31 36 18     && 23 50 47.2     & 31 19 37     && 0.079 && 010131 \\
9CJ2353+2931 && 23 53 58.6     & 29 31 04     && 23 51 26.0     & 29 14 23     && 0.035 && 010329 \\
9CJ2354+2758 && 23 54 48.4     & 27 58 31     && 23 52 15.6     & 27 41 49     && 0.029 && 010315 \\
9CJ2355+3032 && 23 55 03.2     & 30 32 51     && 23 52 30.5     & 30 16 09     && 0.027 && 010130 \\
9CJ2355+3150 && 23 55 20.6     & 31 50 41     && 23 52 47.9     & 31 33 59     && 0.035 && 010131 \\
9CJ2355+2835 && 23 55 54.1     & 28 35 55     && 23 53 21.2     & 28 19 14     && 0.118 && 010329 \\
9CJ2357+2747 && 23 57 36.0     & 27 47 21     && 23 55 02.8     & 27 30 39     && 0.041 && 010119 \\
9CJ2358+3129 && 23 58 13.7     & 31 29 58     && 23 55 40.4     & 31 13 16     && 0.029 && 010209 \\
9CJ2358+2754 && 23 58 34.2     & 27 54 09     && 23 56 00.9     & 27 37 27     && 0.064 && 010119 \\
9CJ2358+2936 && 23 58 47.7     & 29 36 09     && 23 56 14.3     & 29 19 27     && 0.036 && 010131 \\
9CJ2359+2703 && 23 59 00.2     & 27 03 25     && 23 56 26.8     & 26 46 43     && 0.042 && 010315 \\
9CJ2359+3021 && 23 59 19.5     & 30 21 04     && 23 56 46.1     & 30 04 22     && 0.030 && 010124 \\

\hline
\end{tabular}
\end{table*}

\begin{table*}
\caption{Source list for the VSA2 field}
\begin{tabular}{ccccccccccc}
Source name  &&    RA J2000 &  Dec J2000    &&    RA B1950 & Dec B1950    &&   Flux &&  Date  \\
&&&&&&&&density (Jy) \\
\hline
\\
9CJ0915+2933 && 09 15 52.5     & 29 33 21     && 09 12 53.6     & 29 45 52     && 0.148 && 000816 \\
9CJ0917+3446 && 09 17 16.2     & 34 46 36     && 09 14 11.8     & 34 59 11     && 0.067 && 001225 \\
9CJ0919+3324 && 09 19 08.8     & 33 24 42     && 09 16 06.2     & 33 37 23     && 0.524 && 000307 \\
9CJ0920+3312 && 09 20 21.0     & 33 12 39     && 09 17 18.9     & 33 25 23     && 0.058 && 000307 \\
9CJ0920+2755 && 09 20 27.5     & 27 55 52     && 09 17 30.8     & 28 08 37     && 0.052 && 000212 \\
9CJ0921+2834 && 09 21 52.9     & 28 34 53     && 09 18 55.7     & 28 47 42     && 0.057 && 000107 \\
9CJ0922+3232 && 09 22 26.0     & 32 32 26     && 09 19 24.9     & 32 45 16     && 0.025 && 000213 \\
9CJ0923+3107 && 09 23 48.0     & 31 07 56     && 09 20 48.5     & 31 20 50     && 0.148 && 000220 \\
9CJ0923+2815 && 09 23 51.6     & 28 15 26     && 09 20 54.9     & 28 28 20     && 0.374 && 000107 \\
9CJ0925+3026 && 09 25 25.3     & 30 26 35     && 09 22 26.7     & 30 39 33     && 0.025 && 000218 \\
9CJ0925+3159 && 09 25 32.8     & 31 59 54     && 09 22 32.6     & 32 12 53     && 0.068 && 000213 \\
9CJ0925+3626 && 09 25 39.2     & 36 26 30     && 09 22 34.3     & 36 39 29     && 0.040 && 010220 \\
9CJ0925+3127 && 09 25 43.6     & 31 27 12     && 09 22 44.1     & 31 40 11     && 0.232 && 000220 \\
9CJ0925+3612 && 09 25 51.9     & 36 12 38     && 09 22 47.3     & 36 25 37     && 0.265 && 001108 \\
9CJ0926+2758 && 09 26 45.2     & 27 58 22     && 09 23 49.2     & 28 11 24     && 0.062 && 000108 \\
9CJ0926+2711 && 09 26 59.4     & 27 11 06     && 09 24 04.1     & 27 24 09     && 0.030 && 000307 \\
9CJ0927+2954 && 09 27 22.5     & 29 54 13     && 09 24 24.7     & 30 07 17     && 0.029 && 991127 \\
9CJ0927+3034 && 09 27 39.8     & 30 34 16     && 09 24 41.4     & 30 47 20     && 0.063 && 991127 \\
9CJ0927+2738 && 09 27 56.0     & 27 38 18     && 09 25 00.4     & 27 51 23     && 0.026 && 000220 \\
9CJ0928+2904 && 09 28 15.0     & 29 04 16     && 09 25 18.2     & 29 17 22     && 0.027 && 991127 \\
9CJ0930+3601 && 09 30 33.7     & 36 01 15     && 09 27 30.1     & 36 14 27     && 0.210 && 000611 \\
9CJ0930+3503 && 09 30 55.3     & 35 03 38     && 09 27 52.8     & 35 16 51     && 0.340 && 000611 \\
9CJ0931+2734 && 09 31 01.9     & 27 34 50     && 09 28 06.8     & 27 48 04     && 0.035 && 000307 \\
9CJ0931+2750 && 09 31 51.8     & 27 50 52     && 09 28 56.5     & 28 04 08     && 0.087 && 000213 \\
9CJ0932+2837 && 09 32 14.3     & 28 37 31     && 09 29 18.3     & 28 50 48     && 0.058 && 000819 \\
9CJ0932+3339 && 09 32 55.1     & 33 39 29     && 09 29 54.3     & 33 52 48     && 0.124 && 000106 \\
9CJ0933+2845 && 09 33 37.4     & 28 45 32     && 09 30 41.5     & 28 58 52     && 0.036 && 991127 \\
9CJ0933+3254 && 09 33 39.7     & 32 54 42     && 09 30 39.9     & 33 08 02     && 0.029 && 000108 \\
9CJ0934+2756 && 09 34 20.5     & 27 56 04     && 09 31 25.4     & 28 09 26     && 0.028 && 000215 \\
9CJ0934+3050 && 09 34 47.3     & 30 50 57     && 09 31 49.6     & 31 04 21     && 0.045 && 991127 \\
9CJ0935+3633 && 09 35 31.8     & 36 33 18     && 09 32 28.5     & 36 46 44     && 0.149 && 001008 \\
9CJ0935+2917 && 09 35 36.2     & 29 17 11     && 09 32 40.1     & 29 30 37     && 0.036 && 991127 \\
9CJ0936+3207 && 09 36 03.8     & 32 07 14     && 09 33 05.1     & 32 20 40     && 0.041 && 991127 \\
9CJ0936+2554 && 09 36 06.7     & 25 54 10     && 09 33 13.6     & 26 07 37     && 0.029 && 001105 \\
9CJ0936+3313 && 09 36 09.4     & 33 13 09     && 09 33 09.7     & 33 26 36     && 0.037 && 000106 \\
9CJ0936+2624 && 09 36 14.2     & 26 24 04     && 09 33 20.7     & 26 37 31     && 0.063 && 001105 \\
9CJ0937+3206 && 09 37 06.4     & 32 06 56     && 09 34 07.8     & 32 20 25     && 0.059 && 000106 \\
9CJ0937+3411 && 09 37 16.5     & 34 11 33     && 09 34 15.9     & 34 25 03     && 0.062 && 000221 \\
9CJ0938+2611 && 09 38 20.6     & 26 11 33     && 09 35 27.5     & 26 25 06     && 0.031 && 001110 \\
9CJ0938+2559 && 09 38 54.7     & 25 59 40     && 09 36 01.8     & 26 13 15     && 0.037 && 001114 \\
9CJ0939+2908 && 09 39 01.6     & 29 08 29     && 09 36 06.1     & 29 22 04     && 0.109 && 000222 \\
9CJ0939+3556 && 09 39 49.9     & 35 56 14     && 09 36 48.0     & 36 09 51     && 0.115 && 000531 \\
9CJ0940+2626 && 09 40 13.5     & 26 26 57     && 09 37 20.4     & 26 40 35     && 0.062 && 001105 \\
9CJ0940+2547 && 09 40 14.5     & 25 47 10     && 09 37 22.0     & 26 00 48     && 0.032 && 001110 \\
9CJ0940+2603 && 09 40 14.7     & 26 03 30     && 09 37 22.0     & 26 17 07     && 0.448 && 001114 \\
9CJ0940+3015 && 09 40 18.8     & 30 15 09     && 09 37 22.5     & 30 28 46     && 0.066 && 000106 \\
9CJ0941+2547 && 09 41 42.9     & 25 47 22     && 09 38 50.6     & 26 01 04     && 0.030 && 001110 \\
9CJ0941+2728 && 09 41 48.1     & 27 28 38     && 09 38 54.4     & 27 42 20     && 0.158 && 000222 \\
9CJ0941+2722 && 09 41 52.4     & 27 22 18     && 09 38 58.7     & 27 36 00     && 0.045 && 000221 \\
9CJ0942+3309 && 09 42 15.3     & 33 09 30     && 09 39 16.6     & 33 23 13     && 0.061 && 000107 \\
9CJ0942+2626 && 09 42 31.2     & 26 26 57     && 09 39 38.4     & 26 40 40     && 0.038 && 001105 \\
9CJ0942+3344 && 09 42 36.2     & 33 44 38     && 09 39 36.9     & 33 58 21     && 0.054 && 000326 \\
9CJ0942+3737 && 09 42 54.8     & 37 37 36     && 09 39 51.7     & 37 51 20     && 0.039 && 001008 \\
9CJ0943+2833 && 09 43 02.1     & 28 33 57     && 09 40 07.6     & 28 47 41     && 0.025 && 001011 \\
9CJ0943+3614 && 09 43 19.1     & 36 14 52     && 09 40 17.5     & 36 28 37     && 0.210 && 001011 \\
9CJ0943+3300 && 09 43 33.5     & 33 00 14     && 09 40 35.0     & 33 14 00     && 0.041 && 000222 \\
9CJ0944+3347 && 09 44 20.2     & 33 47 56     && 09 41 21.1     & 34 01 44     && 0.039 && 000222 \\
9CJ0944+2554 && 09 44 42.3     & 25 54 42     && 09 41 50.2     & 26 08 30     && 0.120 && 001201 \\
9CJ0945+2729 && 09 45 15.6     & 27 29 11     && 09 42 22.3     & 27 43 01     && 0.092 && 000308 \\
9CJ0945+3656 && 09 45 23.1     & 36 56 02     && 09 42 21.1     & 37 09 52     && 0.029 && 000831 \\

\end{tabular}
\end{table*}

\begin{table*}
\contcaption{}
\begin{tabular}{ccccccccccc}
Source name  &&    RA J2000 &  Dec J2000    &&    RA B1950 & Dec B1950    &&   Flux &&  Date  \\
&&&&&&&&density (Jy) \\
\hline
\\

9CJ0945+2640 && 09 45 31.0     & 26 40 54     && 09 42 38.3     & 26 54 44     && 0.082 && 000308 \\
9CJ0945+3534 && 09 45 38.2     & 35 34 58     && 09 42 37.6     & 35 48 48     && 0.305 && 000821 \\
9CJ0945+2755 && 09 45 56.7     & 27 55 56     && 09 43 03.1     & 28 09 47     && 0.044 && 000222 \\
9CJ0946+3309 && 09 46 10.9     & 33 09 05     && 09 43 12.8     & 33 22 57     && 0.025 && 000222 \\
9CJ0948+3423 && 09 48 38.6     & 34 23 17     && 09 45 39.7     & 34 37 15     && 0.051 && 000326 \\
9CJ0949+2711 && 09 49 10.9     & 27 11 13     && 09 46 18.4     & 27 25 12     && 0.053 && 000408 \\
9CJ0949+2921 && 09 49 24.2     & 29 21 41     && 09 46 29.9     & 29 35 41     && 0.029 && 000222 \\
9CJ0949+2920 && 09 49 48.2     & 29 20 53     && 09 46 53.9     & 29 34 54     && 0.071 && 000222 \\
9CJ0949+3626 && 09 49 53.0     & 36 26 18     && 09 46 52.4     & 36 40 19     && 0.042 && 001010 \\
9CJ0950+2743 && 09 50 33.7     & 27 43 28     && 09 47 40.9     & 27 57 31     && 0.038 && 000222 \\
9CJ0951+3451 && 09 51 11.5     & 34 51 33     && 09 48 12.6     & 35 05 37     && 0.034 && 000603 \\
9CJ0951+3359 && 09 51 45.7     & 33 59 33     && 09 48 47.7     & 34 13 39     && 0.028 && 000603 \\
9CJ0952+2828 && 09 52 06.1     & 28 28 32     && 09 49 12.9     & 28 42 38     && 0.158 && 000222 \\
9CJ0952+3606 && 09 52 26.5     & 36 06 01     && 09 49 26.6     & 36 20 09     && 0.048 && 001010 \\
9CJ0952+3512 && 09 52 32.0     & 35 12 52     && 09 49 33.0     & 35 26 59     && 0.388 && 001010 \\
9CJ0953+3225 && 09 53 28.0     & 32 25 52     && 09 50 31.7     & 32 40 02     && 0.102 && 000222 \\
9CJ0954+3456 && 09 54 06.7     & 34 56 45     && 09 51 08.2     & 35 10 56     && 0.046 && 000603 \\
9CJ0954+3335 && 09 54 26.8     & 33 35 22     && 09 51 29.6     & 33 49 33     && 0.037 && 000109 \\
9CJ0954+3019 && 09 54 27.6     & 30 19 10     && 09 51 33.2     & 30 33 22     && 0.032 && 000222 \\
9CJ0954+2639 && 09 54 39.8     & 26 39 24     && 09 51 48.3     & 26 53 36     && 0.162 && 000324 \\
9CJ0955+3335 && 09 55 37.9     & 33 35 04     && 09 52 40.9     & 33 49 18     && 0.075 && 000408 \\
9CJ0955+3533 && 09 55 47.7     & 35 33 24     && 09 52 49.0     & 35 47 39     && 0.038 && 001225 \\
9CJ0956+3032 && 09 56 37.7     & 30 32 41     && 09 53 43.4     & 30 46 58     && 0.030 && 000317 \\
9CJ0956+2855 && 09 56 40.7     & 28 55 46     && 09 53 47.7     & 29 10 03     && 0.030 && 000324 \\
9CJ0957+3422 && 09 57 46.4     & 34 22 16     && 09 54 49.1     & 34 36 36     && 0.062 && 000818 \\
9CJ0957+3150 && 09 57 51.1     & 31 50 46     && 09 54 56.0     & 32 05 06     && 0.040 && 000901 \\
9CJ0958+3224 && 09 58 20.9     & 32 24 03     && 09 55 25.4     & 32 38 23     && 0.774 && 000818 \\
9CJ0958+3307 && 09 58 27.0     & 33 07 28     && 09 55 30.8     & 33 21 49     && 0.033 && 000408 \\
9CJ0958+2948 && 09 58 58.9     & 29 48 04     && 09 56 05.6     & 30 02 26     && 0.076 && 000317 \\
9CJ1000+2752 && 10 00 07.6     & 27 52 46     && 09 57 15.9     & 28 07 10     && 0.082 && 001015 \\
9CJ1000+3437 && 10 00 27.4     & 34 37 40     && 09 57 30.3     & 34 52 06     && 0.026 && 000824 \\
9CJ1000+2752 && 10 00 29.4     & 27 52 12     && 09 57 37.8     & 28 06 37     && 0.071 && 001015 \\
9CJ1001+2911 && 10 01 10.2     & 29 11 39     && 09 58 17.6     & 29 26 06     && 0.198 && 000818 \\
9CJ1001+3424 && 10 01 11.9     & 34 24 50     && 09 58 15.2     & 34 39 17     && 0.306 && 000825 \\
9CJ1001+2846 && 10 01 46.3     & 28 46 55     && 09 58 54.1     & 29 01 23     && 0.170 && 000818 \\
9CJ1002+3042 && 10 02 33.0     & 30 42 08     && 09 59 39.5     & 30 56 38     && 0.043 && 000824 \\
9CJ1003+2845 && 10 03 13.5     & 28 45 26     && 10 00 21.6     & 28 59 57     && 0.093 && 000818 \\
9CJ1003+3347 && 10 03 26.8     & 33 47 29     && 10 00 30.9     & 34 02 01     && 0.068 && 000828 \\
9CJ1003+3244 && 10 03 57.6     & 32 44 03     && 10 01 02.6     & 32 58 36     && 0.202 && 000824 \\
9CJ1004+3010 && 10 04 11.6     & 30 10 33     && 10 01 18.8     & 30 25 07     && 0.046 && 001107 \\
9CJ1004+3151 && 10 04 32.9     & 31 51 53     && 10 01 38.8     & 32 06 27     && 0.080 && 000824 \\
9CJ1006+3236 && 10 06 07.7     & 32 36 28     && 10 03 13.2     & 32 51 05     && 0.074 && 010208 \\
9CJ1007+3003 && 10 07 14.9     & 30 03 52     && 10 04 22.5     & 30 18 32     && 0.079 && 010131 \\

\hline
\end{tabular}
\end{table*}

\begin{table*}
\caption{Source list for the VSA3 field}
\begin{tabular}{ccccccccccc}
Source name  &&    RA J2000 &  Dec J2000    &&    RA B1950 & Dec B1950    &&   Flux &&  Date  \\
&&&&&&&&density (Jy) \\
\hline
\\
9CJ1510+4221 && 15 10 17.9     & 42 21 54     && 15 08 29.0     & 42 33 13     && 0.065 && 000622 \\
9CJ1511+4430 && 15 11 42.7     & 44 30 45     && 15 09 57.7     & 44 41 59     && 0.060 && 000815 \\
9CJ1512+4540 && 15 12 27.1     & 45 40 26     && 15 10 44.2     & 45 51 37     && 0.030 && 000622 \\
9CJ1513+4554 && 15 13 50.7     & 45 54 23     && 15 12 08.5     & 46 05 30     && 0.038 && 000622 \\
9CJ1516+4349 && 15 16 31.5     & 43 49 49     && 15 14 46.2     & 44 00 48     && 0.027 && 000816 \\
9CJ1516+4159 && 15 16 59.6     & 41 59 34     && 15 15 11.3     & 42 10 31     && 0.044 && 001106 \\
9CJ1518+4618 && 15 18 44.7     & 46 18 56     && 15 17 04.2     & 46 29 46     && 0.054 && 000816 \\
9CJ1518+4131 && 15 18 47.2     & 41 31 36     && 15 16 58.6     & 41 42 27     && 0.040 && 001108 \\
9CJ1519+4254 && 15 19 27.0     & 42 54 08     && 15 17 40.7     & 43 04 57     && 0.047 && 000816 \\
9CJ1520+4211 && 15 20 39.7     & 42 11 13     && 15 18 52.4     & 42 21 57     && 0.082 && 001106 \\
9CJ1521+4336 && 15 21 49.6     & 43 36 39     && 15 20 04.9     & 43 47 20     && 0.605 && 000816 \\
9CJ1523+4156 && 15 23 09.3     & 41 56 25     && 15 21 22.1     & 42 07 02     && 0.061 && 000301 \\
9CJ1525+4201 && 15 25 23.6     & 42 01 18     && 15 23 36.9     & 42 11 46     && 0.052 && 000301 \\
9CJ1526+4201 && 15 26 45.4     & 42 01 42     && 15 24 58.8     & 42 12 06     && 0.070 && 000301 \\
9CJ1528+4219 && 15 28 00.2     & 42 19 14     && 15 26 14.3     & 42 29 34     && 0.046 && 000316 \\
9CJ1528+4233 && 15 28 19.9     & 42 33 36     && 15 26 34.5     & 42 43 54     && 0.032 && 000301 \\
9CJ1528+4522 && 15 28 41.1     & 45 22 16     && 15 27 00.8     & 45 32 33     && 0.046 && 000222 \\
9CJ1529+4538 && 15 29 10.4     & 45 38 22     && 15 27 30.7     & 45 48 38     && 0.029 && 000223 \\
9CJ1529+3945 && 15 29 49.8     & 39 45 10     && 15 28 00.1     & 39 55 24     && 0.036 && 010126 \\
9CJ1531+4356 && 15 31 02.6     & 43 56 38     && 15 29 20.1     & 44 06 47     && 0.031 && 000222 \\
9CJ1531+4048 && 15 31 40.9     & 40 48 26     && 15 29 53.1     & 40 58 34     && 0.040 && 000316 \\
9CJ1532+4604 && 15 32 50.7     & 46 04 47     && 15 31 12.5     & 46 14 50     && 0.062 && 000222 \\
9CJ1533+4107 && 15 33 27.9     & 41 07 23     && 15 31 40.9     & 41 17 24     && 0.030 && 000316 \\
9CJ1533+3934 && 15 33 44.3     & 39 34 20     && 15 31 54.9     & 39 44 20     && 0.032 && 010126 \\
9CJ1534+3834 && 15 34 50.5     & 38 34 10     && 15 32 59.7     & 38 44 06     && 0.026 && 010402 \\
9CJ1536+3833 && 15 36 13.8     & 38 33 28     && 15 34 23.2     & 38 43 19     && 0.183 && 010402 \\
9CJ1536+4627 && 15 36 17.3     & 46 27 33     && 15 34 40.4     & 46 37 23     && 0.037 && 000223 \\
9CJ1536+3845 && 15 36 23.2     & 38 45 52     && 15 34 32.9     & 38 55 43     && 0.107 && 010402 \\
9CJ1538+4225 && 15 38 55.8     & 42 25 27     && 15 37 11.8     & 42 35 09     && 0.055 && 000223 \\
9CJ1539+4217 && 15 39 25.6     & 42 17 27     && 15 37 41.6     & 42 27 07     && 0.031 && 000223 \\
9CJ1539+4735 && 15 39 34.8     & 47 35 31     && 15 38 00.9     & 47 45 10     && 0.191 && 000604 \\
9CJ1539+4602 && 15 39 35.1     & 46 02 49     && 15 37 58.1     & 46 12 28     && 0.027 && 000223 \\
9CJ1540+4138 && 15 40 43.0     & 41 38 18     && 15 38 58.0     & 41 47 53     && 0.030 && 000223 \\
9CJ1541+4114 && 15 41 01.3     & 41 14 28     && 15 39 15.6     & 41 24 02     && 0.035 && 000604 \\
9CJ1541+4456 && 15 41 10.2     & 44 56 31     && 15 39 31.3     & 45 06 05     && 0.035 && 000223 \\
9CJ1541+4727 && 15 41 44.3     & 47 27 51     && 15 40 10.5     & 47 37 22     && 0.027 && 000823 \\
9CJ1542+4359 && 15 42 22.8     & 43 59 13     && 15 40 42.2     & 44 08 42     && 0.027 && 000820 \\
9CJ1545+4751 && 15 45 08.6     & 47 51 55     && 15 43 36.2     & 48 01 13     && 0.190 && 000820 \\
9CJ1545+4130 && 15 45 21.4     & 41 30 27     && 15 43 36.8     & 41 39 46     && 0.058 && 000817 \\
9CJ1545+4622 && 15 45 25.5     & 46 22 45     && 15 43 50.1     & 46 32 03     && 0.026 && 000823 \\
9CJ1546+4257 && 15 46 22.6     & 42 57 57     && 15 44 40.7     & 43 07 11     && 0.038 && 000817 \\
9CJ1547+4208 && 15 47 59.0     & 42 08 53     && 15 46 16.0     & 42 18 01     && 0.076 && 000824 \\
9CJ1548+4031 && 15 48 11.3     & 40 31 27     && 15 46 25.5     & 40 40 35     && 0.070 && 000824 \\
9CJ1550+4536 && 15 50 43.8     & 45 36 24     && 15 49 07.7     & 45 45 23     && 0.031 && 000825 \\
9CJ1550+4545 && 15 50 54.8     & 45 45 28     && 15 49 19.0     & 45 54 25     && 0.027 && 000825 \\
9CJ1553+4039 && 15 53 15.6     & 40 39 27     && 15 51 30.6     & 40 48 16     && 0.036 && 001118 \\
9CJ1554+4350 && 15 54 15.7     & 43 50 29     && 15 52 36.6     & 43 59 14     && 0.036 && 010629 \\
9CJ1554+4348 && 15 54 42.3     & 43 48 19     && 15 53 03.3     & 43 57 03     && 0.042 && 000828 \\
9CJ1556+4257 && 15 56 36.2     & 42 57 07     && 15 54 55.8     & 43 05 44     && 0.115 && 000827 \\
9CJ1556+4259 && 15 56 57.7     & 42 59 42     && 15 55 17.4     & 43 08 18     && 0.056 && 000827 \\
9CJ1557+4522 && 15 57 19.0     & 45 22 21     && 15 55 43.4     & 45 30 55     && 0.145 && 001118 \\
9CJ1558+4146 && 15 58 24.1     & 41 46 36     && 15 56 41.8     & 41 55 06     && 0.029 && 000827 \\
9CJ1559+4349 && 15 59 31.3     & 43 49 16     && 15 57 53.0     & 43 57 42     && 0.059 && 000828 \\
9CJ1601+4123 && 16 01 27.9     & 41 23 51     && 15 59 45.3     & 41 32 09     && 0.066 && 010621 \\
9CJ1601+4316 && 16 01 40.5     & 43 16 48     && 16 00 01.5     & 43 25 06     && 0.042 && 001117 \\

\end{tabular}
\end{table*}

\label{lastpage}

\end{document}